\documentstyle[epsfig,sprocl]{article}
\input{epsf.tex}
\font \fwork = cmssqi8

\newcommand{\be}{\begin{equation}}
\newcommand{\ee}{\end{equation}}
\newcommand{\ba}{\begin{eqnarray}}
\newcommand{\ea}{\end{eqnarray}}

\newcommand{\nsigma}{\mbox{\boldmath $\sigma$}}

\newcommand{\np}{{\bf      p}}       
\newcommand{\nq}{{\bf      q}}

\newcommand{\nP}{{\bf      P}}

\begin{document}
\hspace*{-18pt}{\epsfxsize=80pt  \epsfbox{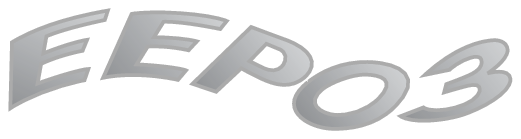}} 
\vspace*{-16pt}
\rightline{\fwork LPSC Grenoble, France, October 14-17, 2003}
\vspace*{15pt}
%
%
\title{\Large\bf ELECTROMAGNETIC (${\vec e},e'{\vec p}$) OBSERVABLES \& RELATIVISTIC NUCLEON DYNAMICS}

\author{\underline{J.A. Caballero$^1$}, T.W. Donnelly$^2$, M.C. Mart\'{\i}nez$^1$, E. Moya de Guerra$^3$,
J.M. Ud\'{\i}as$^4$, J.R. Vignote$^4$}

\address{
    {\it $^1$Departamento de F\'{\i}sica At\'omica, Molecular y Nuclear,
             Universidad de Sevilla, Apdo. 1065, E-41080 Sevilla, SPAIN \\
         $^2$Center for Theoretical Physics, Laboratory for Nuclear Science
             and Department of Physics, Massachusetts Institute of Technology, Cambridge, 
             MA 02139, USA \\
         $^3$Instituto de Estructura de la Materia, CSIC, Serrano 123, E-28006 Madrid, SPAIN \\
         $^4$Departamento de F\'{\i}sica At\'omica, Molecular y Nuclear, Universidad Complutense de
             Madrid, E-28040 Madrid, SPAIN}
        }
\maketitle
%
%
At present there exists a great interest in the search for evidence of possible modification of 
the nucleon form factors inside the nuclear medium. 
Recent theoretical work~\cite{Lu98} predict changes 
in the form factors within the experimental limits. Importantly, the longitudinal to sideways 
transferred polarization ratio has been identified as being ideally suited for such studies, 
as these polarization observables are believed to be the least
sensitive to most standard nuclear structure uncertainties while their ratio shows a high 
sensitivity to the ratio of the electric to magnetic form factors. 
The kinematic regime where the measurements have been undertaken is at relatively high energy and
it is clear that relativistic effects in wave functions and operators are essential. 
In the relativistic distorted wave impulse approximation (RDWIA), nucleon wave functions are
described by solutions of the Dirac equation with scalar and vector (S-V) potentials, and the
relativistic free nucleon current operator is used. 
So far, RDWIA calculations for cross sections and response functions at low and high missing 
momenta~\cite{Udi93} have clearly improved 
the comparison with experimental data over the previous non-relativistic approaches.
In this work we focus on the analysis of polarized 
$^{16}O(\vec{e},e'\vec{p})^{15}N$ observables.
Our aim is to explore a selected set of model dependences that could contaminate any attempt 
to infer medium modifications, mainly related to the description of FSI and to the role played by 
relativity.

In what follows we briefly review the general formalism needed to describe coincidence 
$(\vec{e},e'\vec{p})$ reactions. We consider plane waves for the incoming and outgoing electron 
(treated in the extreme relativistic limit) and the Born approximation (one virtual photon exchanged). 
When the incoming electron is polarized and the final nucleon polarization is measured,
the differential cross section can be written as~\cite{Bof96,Pick87}
\begin{equation}
\frac{d\sigma}{d\varepsilon_ed\Omega_ed\Omega_F}=\frac{\sigma_0}{2}[1+\nP\cdot
\nsigma+h(A+\nP'\cdot\nsigma)]\, ,
\label{difcross1}
\end{equation}
where the variables $\{\varepsilon_e, \Omega_e\}$ refer to the scattered electron and 
$\Omega_F$ to the ejected nucleon. The term
$\sigma_0$ is the unpolarized cross section, $h$ is the incident
electron helicity, $A$ denotes the electron analyzing power, and
$\nP$ ($\nP'$) represents the induced (transferred) polarization. 
Here we limit our attention to the longitudinal and sideways transferred polarization asymmetries 
$P'_{l}$, $P'_{s}$. 

Within RDWIA, the nucleon current matrix elements needed to compute each observable is calculated as
\begin{equation}
J^{\mu}_N(\omega,\nq)=\int{d\np\overline{\Psi}_F(\np+\nq)\hat{J}^{\mu}_N\Psi_B(\np)}\, ,
\label{relcurrent}
\end{equation}
where $\Psi_B$ and $\Psi_F$ are 
relativistic wave functions describing the initial bound and final outgoing nucleons, respectively, and
$\hat{J}^{\mu}_N$ is the relativistic one-body current operator.
The bound wave function $\Psi_B$ is a four-spinor with well-defined parity and angular momentum quantum 
numbers $\kappa_b$, $\mu_b$, obtained within the framework of the relativistic independent particle 
shell model. 
The wave function for the ejected proton $\Psi_F$ is a scattering solution of a Dirac-like equation, 
which includes S-V global optical potentials obtained by fitting elastic proton scattering data. 
Finally, for the nucleon current operator we consider the two usual 
choices denoted as CC1 and CC2~\cite{For83}.

As is well known, the presence of the S-V potentials leads to a significant
dynamical enhancement of the lower components of the Dirac solution at the nuclear interior.
This effect has been also referred to as
{\it spinor distortion}~\cite{Kel97}.
The analysis of these dynamical effects can be done by constructing properly normalized four-spinor
wave functions where the negative-energy components have been projected out. Thus, instead of the fully
relativistic expression given in Eq.~(\ref{relcurrent}), the nucleon current is evaluated as
\begin{equation}
J^{\mu (+,+)}_N(\omega,\nq)=\int{d\np\overline{\Psi}_F^{(+)}(\np+\nq)\hat{J}^{\mu}_N\Psi_B^{(+)}(\np)}
\, ,
\label{projcurrent}
\end{equation}
where $\Psi_B^{(+)}(\np)$, ($\Psi_F^{(+)}(\np)$) is the positive-energy projection of 
$\Psi_B(\np)$, ($\Psi_F(\np)$).
Notice that the relationship between lower and upper components 
in the projected wave functions is similar to that corresponding to
free nucleon wave functions, but with the positive-energy projectors depending explicitly on the
integration variable $\np$. An additional approach, referred to as
asymptotic projection, consists of introducing the asymptotic values of the momenta into the positive-energy
projectors. This
asymptotic projection is very similar to the effective
momentum approximation (EMA-noSV) introduced originally by Kelly~\cite{Kel97}, where
the four spinors used have 
the same upper components as those of the Dirac equation solutions, but the lower components are 
obtained by enforcing the ``free'' relationship between upper and lower components and using
the asymptotic momenta at the nucleon vertex. 

We analyze the recoil nucleon transferred
polarization asymmetries for proton knockout from $^{16}O$.
Coulomb gauge has been assumed and 
the bound nucleon wave function is obtained 
using the parameters of the set NLSH~\cite{Sharma93}. Results computed with other parameterizations
are found to be similar.
For the outgoing nucleon wave function, we show and compare results using different relativistic 
optical potentials parametrizations~\cite{Coo93}, which provide the best
phenomenological global optical potentials available in the literature. 
Finally, the nucleon form factor
parameterization of Gari and Krumplemann~\cite{Gar85} is considered.

\begin{figure}
{\par\centering \resizebox*{0.95\textwidth}{0.25\textheight}{\rotatebox{270}
{\includegraphics{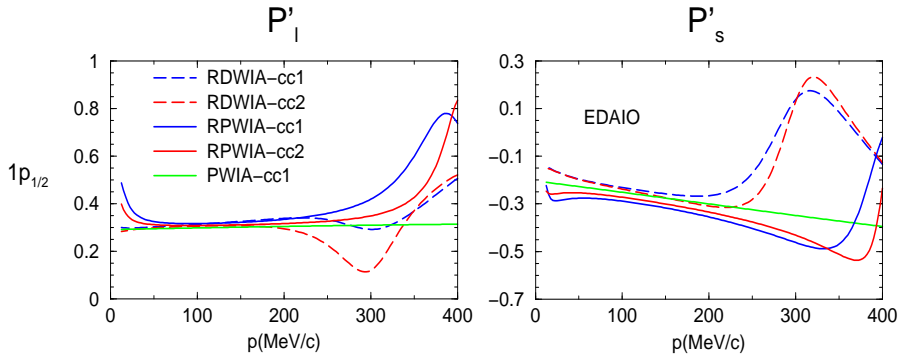}}} \par}
\caption{Longitudinal and sideways transferred polarization asymmetries for proton knockout from
the $1p_{1/2}$ shell in $^{16}O$ (see text and labels for details). \label{fig2}}
\end{figure}

The kinematics is chosen with 
$(q,\omega)$ constant,  $q=1$ GeV/c and $\omega=439$ MeV, yielding $|Q^2|=0.8$ (GeV/c)$^2$.  
This roughly corresponds to the experimental 
conditions of experiments E89-003 and E89-033 performed at JLab~\cite{Gao}. 
In Fig.~\ref{fig2} we compare the results within the relativistic plane wave impulse
approximation (RPWIA), that includes the dynamical enhancement of the 
lower components of the bound nucleon wave function but neglects the distortion 
effects in the scattered wave function, with plane wave calculations after projecting out
these negative-energy components, denoted as PWIA. As explained 
in~\cite{Cris1}, the RPWIA and their respective positive-energy projection results 
mainly differ for $p\ge 250-300$ MeV/c, although there are sizeable differences even for 
low/medium $p$ values, in particular for $P'_s$.
More importantly, RPWIA results present oscillations
that are completely absent in PWIA.
This is connected with the factorization property
which is exactly recovered when projecting out the negative energy components.
RDWIA results evaluated using the
energy-dependent, A-independent optical potential derived for $^{16}O$ (EDAIO),
are also presented in Fig.~\ref{fig2}.
For low missing momentum values $p\leq 200$ MeV/c, the effects of FSI do not modify substantially
the behaviour of the polarization asymmetries, particularly when compared with PWIA.
Similar comments also apply to
the $p_{3/2}$ and $s_{1/2}$ shells, although in these cases a smaller effect is observed
for $P'_s$. It is important to point out that FSI lead to a significant reduction
of the individual response functions.
Hence, the results in Fig.~\ref{fig2} clearly
indicate that for low $p$-values, FSI effects are partially cancelled when constructing the transferred
polarization asymmetries.
For high missing momentum, $p\geq 200$ MeV/c, FSI strongly modify the behaviour of the polarizations.
The oscillations appearing in RPWIA results are also present in RDWIA,
although the maxima and minima are located at 
different $p$-values. 

\begin{figure}
{\par\centering \resizebox*{0.95\textwidth}{0.25\textheight}{\rotatebox{270}
{\includegraphics{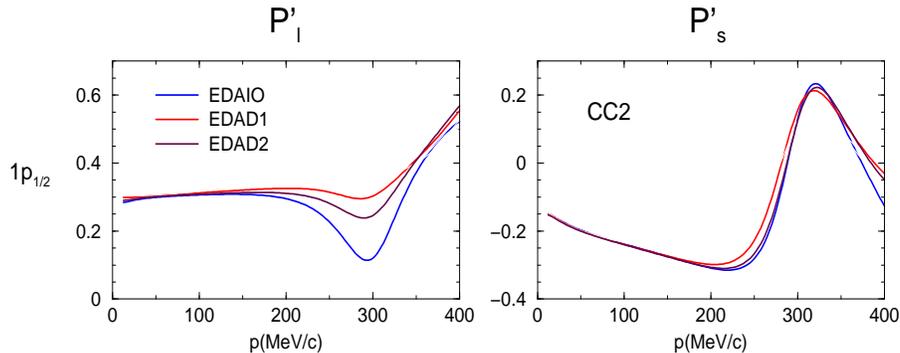}}} \par}
\caption{Same observables as in Fig.~\ref{fig2}. All the results correspond to RDWIA with the CC2 current, 
but using different optical potentials (see labels). \label{fig3}}
\end{figure}

Next we focus on the uncertainties introduced by different 
relativistic optical 
potentials. In Fig.~\ref{fig3} we present $P'_l$ and $P'_s$ for the
$p_{1/2}$ shell evaluated using three different relativistic optical potential 
parameterizations: EDAIO, EDAD1 and EDAD2~\cite{Coo93}. 
Transferred polarization asymmetries 
are expected to be relatively insensitive to the choice of optical potential at low missing momenta. 
This can be seen in Fig.~\ref{fig2}, at least up to $p= 150$ MeV/c.
For larger $p$-values, $P'_l$ exhibits a strong dependence on the optical potential.
Note also that the current operator 
choice gives rise to very significant differences in $P'_l$ within this $p$-region, 
being of the same order as those introduced
by the optical potentials. In the case of $P'_s$, 
in general less dependence on the interaction model as well as on the current is observed.
We conclude that both transferred polarization
asymmetries at moderate $p$ values ($p \simeq 100$ MeV/c) are independent 
of the optical potential choice. Increasing $p$ from here, 
each optical potential corresponds to a different curve especially for $P'_l$.
However, caution should be placed on this 
kinematical region because other ingredients beyond 
the impulse approximation, such as meson exchange currents (MEC), 
$\Delta$-isobar, short-range correlations, {\it etc.,} may also play a crucial role.

\begin{figure}
{\par\centering \resizebox*{0.95\textwidth}{0.25\textheight}{\rotatebox{270}
{\includegraphics{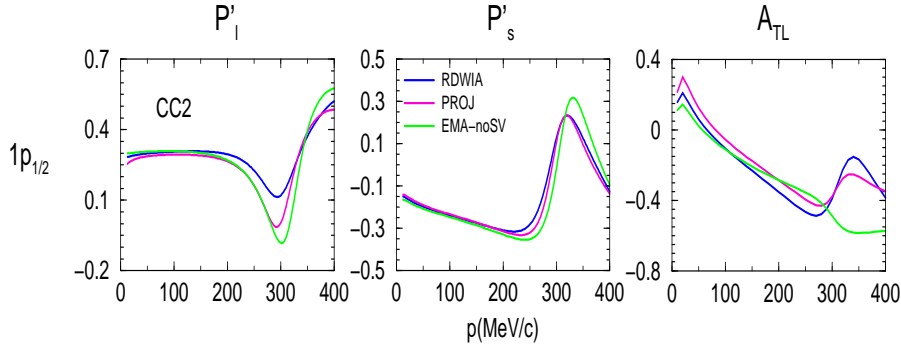}}} \par}
\caption{$P'_l$  and $P'_s$ (left and middle panels, respectively) and $A_{TL}$ (right panel) for the $1p_{1/2}$ shell in $^{16}O$ (see text for details). All results correspond to CC2 current choice.\label{fig4}}
\end{figure}

In Fig.~\ref{fig4} we focus on dynamical relativistic effects.
All of the results have been obtained using EDAIO~\cite{Coo93}. To make explicit the effects 
introduced by
spinor distortion, in each graph we compare the fully relativistic calculations (blue lines) with 
the results after projecting out the negative-energy components (pink lines). Finally we 
also present for reference the results corresponding to the
EMA-noSV approach (green line). 
In contrast to the plane wave limit, results
in Fig.~\ref{fig4} make it clear that, within the relativistic distorted wave approximation, the 
oscillatory behaviour persists even after projecting the bound and scattered
proton wave functions over positive-energy states. The same comment
applies to the EMA-noSV approach. On the contrary, this last fact does not apply to 
the left-right asymmetry $A_{TL}$, as we can see in the right panel of Fig.~\ref{fig4}. 
Note also that model dependences at low $p$ are substantially larger for $A_{TL}$.

\begin{figure}[t]
{\par\centering \resizebox*{0.8\textwidth}{0.45\textheight}{\rotatebox{270}
{\includegraphics{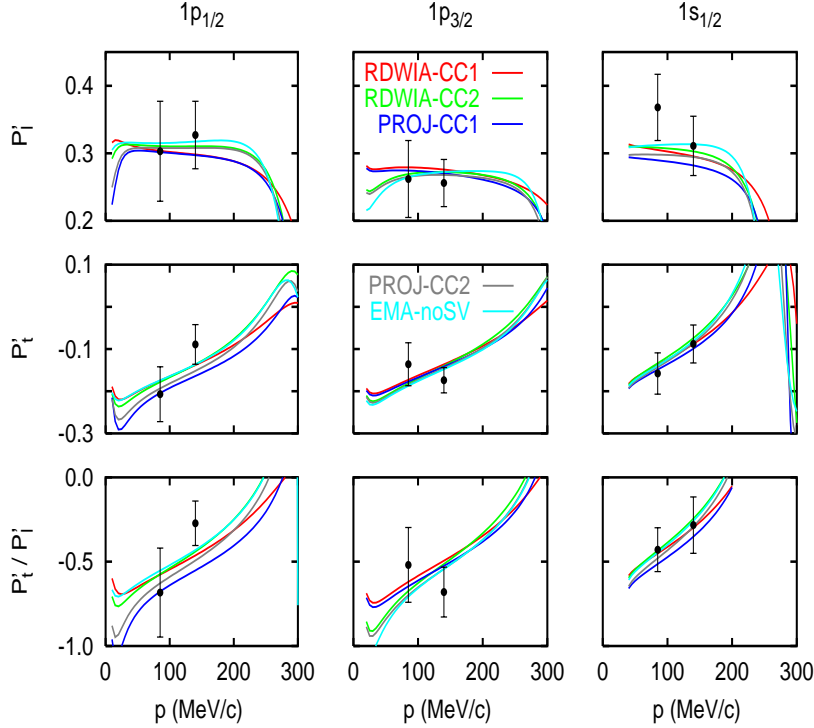}}} \par}
\caption{$P'_l$, $P'_t$ and the ratio $P'_t/P'_l$ compared with experimental
data (see text). \label{fig5}}
\end{figure}

Finally, we compare our calculations with the experimental data measured
at JLab~\cite{Mal00}. 
Fig.~\ref{fig5} shows $P'_l$ (top panels), $P'_t$ (middle panels)
and the ratio $P'_t/P'_l$ (bottom panels) for proton knockout in $^{16}O$ from the
$1p_{1/2}$ (left panels), $1p_{3/2}$ (middle panels) and $1s_{1/2}$ (right panels) shells. 
Note the change of notation for the transverse polarization ratio~\cite{Cris3}.
Curves corresponding to RDWIA, positive-energy projected and EMA-noSV calculations are presented. 
All of the results have been obtained using the EDAIO potential.
As shown, all theoretical calculations satisfactorily reproduce the data, 
improving somehow the general agreement compared with previous 
semi-relativistic (SR) analyses~\cite{Far03}. 

Summarizing, an analysis of recoil nucleon polarized $(\vec{e},e'\vec{p})$
asymmetries within RDWIA has been performed in the case
of proton knockout from $^{16}O$ and quasi-perpendicular kinematics.
Effects linked to the description of FSI and dynamical relativity 
have been studied. FSI constitutes a basic ingredient in order to get 
reliable results, and our model presents a great stability
for low missing momenta. Also effects linked to relativity are shown to 
be quite modest in this $p$-region. We conclude that this region may allow to look for 
other more exotic effects such as medium modifications of the nucleon form factors. On the 
other hand, although being aware of the experimental difficulties, measurements at higher $p$ 
could provide a good test of the different model ingredients.
Other effects that go
beyond the impulse approximation, such as meson exchange
currents and the $\Delta$-isobar contribution,
may also play a very important role. These remain to be investigated in
a relativistic context.
In the final analysis, any interpretation in terms of medium modified 
nucleon form factors requires having excellent control of all of these  
model dependences.


\end{document}